\documentclass[prl,preprint,superscriptaddress, amsfonts,amssymb,amsmath,titlepage]{revtex4-1}
\usepackage[utf8]{inputenc}
\usepackage[hidelinks, breaklinks=true]{hyperref}
\usepackage{graphicx}
\usepackage{textcomp}
\usepackage{xcolor}
\usepackage{verbatim}
\usepackage{xspace} % Necessary to manage a nice 'space' at the end of \newcommand
\usepackage{ulem} % Use then \sout{} to cross a word out
\usepackage{setspace} % Needed to modify line spacing (presently used for caption of Fig. 1)

\newcommand{\etal}{et al.\xspace}
\newcommand{\myref}[7]{\href{http://dx.doi.org/#7}{#1 #2. \textsl{#3} \textbf{#4}, #5 (#6).}} %Authors, title, Journal, vol, page, year, doi

\newcommand{\PTCDAmatrix}{$\left(\protect\smallmatrix ~0.6&~5.2 \\ -8.2&~0.8 \endsmallmatrix \right)$\xspace}
\newcommand{\Omatrix}{$\left(\protect\smallmatrix ~2&~2 \\ -1&~1 \endsmallmatrix \right)$\xspace}
\newcommand{\OWoodN}{$(\sqrt{2} \times 2 \sqrt{2})$R$45^\circ$-$2$O\xspace}

\newcommand{\pgi}{Peter Grünberg Institut (PGI-3), Forschungszentrum Jülich, 52425 Jülich, Germany}
\newcommand{\jara}{Jülich Aachen Research Alliance (JARA), Fundamentals of Future Information Technology, 52425 Jülich, Germany}
\newcommand{\rwth}{Experimentalphysik IV A, RWTH Aachen University, Otto-Blumenthal-Straße, 52074 Aachen, Germany}
\newcommand{\marburg}{Fachbereich Physik und Zentrum für Materialwissenschaften, Philipps-Universität Marburg, Renthof 5, 35032 Marburg, Germany}
\newcommand{\graz}{Institute of Physics, University of Graz, NAWI Graz, 8010 Graz, Austria}

\begin{document}

\title{Tracing orbital images on ultrafast time scales}

\author{R.~Wallauer*}    \affiliation{\marburg}
\author{M.~Raths}       \affiliation{\pgi} \affiliation{\jara} \affiliation{\rwth}
\author{K.~Stallberg}   \affiliation{\marburg}
\author{L.~Münster}     \affiliation{\marburg}
\author{D.~Brandstetter}\affiliation{\graz}
\author{X.~Yang}         \affiliation{\pgi} \affiliation{\jara} \affiliation{\rwth}
\author{J.~Güdde}       \affiliation{\marburg}
\author{P.~Puschnig}    \affiliation{\graz}
\author{S.~Soubatch}    \affiliation{\pgi} \affiliation{\jara}
\author{C.~Kumpf}       \affiliation{\pgi} \affiliation{\jara} \affiliation{\rwth}
\author{F.~C.~Bocquet}  \affiliation{\pgi} \affiliation{\jara}
\author{F.~S.~Tautz*}    \affiliation{\pgi} \affiliation{\jara} \affiliation{\rwth}
\author{U.~Höfer}       \affiliation{\marburg}

\date{\today}

\maketitle

\textbf{Frontier orbitals, i.e., the highest occupied and lowest unoccupied orbitals of a molecule, generally determine molecular properties, such as chemical bonding and reactivities \cite{Woodward1965, Fukui1982}. Consequently, there has been a lot of  interest in measuring them, despite the fact that, strictly speaking, they are not quantum-mechanical observables \cite{Itatani2004, Repp2005, Meckel2008, Cocker2016, Truhlar2019}.  Yet, with photoemission tomography a powerful technique has recently been introduced by which the electron distribution in orbitals of molecules adsorbed at surfaces can be imaged in momentum space \cite{Puschnig2009}. This has even been used for the identification of reaction intermediates in surface reactions \cite{Yang2019}. However, so far it has been impossible to follow an orbital's momentum-space dynamics in time, for example through an excitation process or a chemical reaction. Here, we report a key step in this direction: we combine time-resolved photoemission employing high laser harmonics \cite{Rohwer2011, Heyl2012, Eich2017, Nicholson2018, Gierz2013, Na2019} 
and a recently developed momentum microscope \cite{Tusche2016} to establish a tomographic, femtosecond pump-probe experiment of unoccupied molecular orbitals. Specifically, we measure the full momentum-space distribution of transiently excited electrons. Because in molecules this momentum-space distribution is closely linked to orbital shapes, our experiment offers the extraordinary possibility to observe ultrafast electron motion in time and space. This enables us to connect their excited states dynamics to specific real-space excitation pathways.
}

The basis for a quantum-mechanical description of matter is the many-electron wave function. At various levels of approximation, up to the exact configuration interaction wave function, one can write it in terms of single-electron wave functions: the orbitals. Their great advantage is that they provide a link between spectral properties and spatial electron distributions, which is of obvious benefit in chemistry \cite{Woodward1965, Fukui1982}. But in spite of being such a powerful concept, orbitals have their subtleties. As complex entities with amplitude and phase they are not quantum mechanical observables and also, from a fundamental point of view, they do not describe the behaviour of individual electrons prior to removing them from the many-electron system \cite{Truhlar2019}.

In recent years, however, it has been shown that a wealth of important information can be extracted from orbitals as accessed experimentally by electron removal (or addition) spectroscopies. For example, the frontier orbitals of $\pi$-conjugated molecules derived from photoemission tomography \cite{Puschnig2009} have been used to deconvolve electronic spectra of molecules in momentum space beyond energy resolution limits \cite{Puschnig2016}, to determine molecular orientations \cite{Kliuiev2019}, and to pinpoint charge transfer \cite{Ziroff2010}. Photoemission tomography can even be used to reconstruct the spatial probability amplitudes of orbitals in two and three dimensions \cite{Puschnig2009, Wiessner2014, Lueftner2014, Weiss2015}. The best way to think about such as-measured orbitals is to interpret them as Dyson orbitals that quantify the change between the $N$ electron wave function before and the $N-1$ electron wave function after photoionisation \cite{Dauth2011, Ortiz2020}. If Koopman's theorem \cite{Koopmans1934} holds, which should be the case for molecules with weakly correlated electrons, this change is closely related to canonical Hartree-Fock or Kohn-Sham orbitals \cite{Truhlar2019}. As it turns out, this is the situation that we encounter here.

Addressing molecular orbitals of excited states in pump-probe experiments would bring photoemission tomography to its culmination, because it would empower this technique to provide the coveted access to molecular excitation and reaction dynamics in time \textit{and space}: instead of just recording the photoemission intensity from the corresponding energy level, the evolution of the $N$ electron wave function after excitation could be traced by  monitoring the complete Dyson orbital in momentum space.

However, this extension meets with difficulties. First, the most prominent features in frontier orbital photoemission from organic molecules appear at $\sim 1.4-1.7$~\AA$^{-1}$ momentum parallel to the surface \cite{Puschnig2009}. This reflects the periodicity of the molecular C--C bond network. For photoemission with conventional laser sources, such high parallel momenta are out of reach. Recently, however, high enough probe photon energies have become available through laser high-harmonic generation (HHG) and have enabled time-resolved photoemission experiments to record band structure movies of solids, i.e., to trace the temporal evolution of the electron system over the complete Brillouin zone  \cite{Rohwer2011, Heyl2012, Eich2017, Nicholson2018, Gierz2013, Na2019}. Here, we employ such a HHG light source \cite{Heyl2012}, combine it with tunable pump pulses for resonant excitation and probe the photoelectron distribution with a momentum microscope \cite{Tusche2016} (see Methods).

Photoelectron tomography requires both a conductive and a sufficiently corrugated substrate. The latter promotes a small number of well-defined azimuthal orientations of the molecular adsorbate on the substrate surface, which simplifies the interpretation of tomographic data. If photoelectron tomography is applied to excited states, electronic decoupling of the molecule from the substrate arises as an additional requirement. Here, we employ an ultrathin oxide layer to decouple 3,4,9,10-perylene-tetracarboxylic-dianhydride (PTCDA) molecules from the metallic Cu(001) substrate surface \cite{Yang2018}. At the same time, the submonolayer oxygen coverage provides a \OWoodN surface corrugation that imposes two clearly defined azimuthal orientations of PTCDA on Cu(001)-2O (see the structural model in Fig.~\ref{Fig:ExpOverview}a). For more details, see Methods and Extended Data Fig.~\ref{supp:LEED}.
\newline

\noindent{\bf Time-resolved photoemission tomography}

Fig.~\ref{Fig:ExpOverview}a displays the scheme of our experiment (see Methods for more details). We excite the molecules with 2.3 eV pump pulses and employ 21.7 eV probe pulses at variable delay times $t_p$ for photoemission. In our momentum microscope, the parallel photoelectron momenta $k_x$ and $k_y$ are mapped onto the detector, while the energy $E$ is retrieved simultaneously by a time-of-flight measurement. The recorded four-dimensional data cube $I(E,k_x,k_y,t_p)$, where $I$ denotes the photoemission intensity, enables us to deduce the spatial electron distribution in terms of orbitals, their energy position and their time evolution.

Fig.~\ref{Fig:ExpOverview}b shows a cut through such a data cube, displaying $I(E,k_x)$ at zero delay time. As expected for $\pi$-orbitals of flat lying PTCDA molecules, photoemission predominantly occurs at large parallel momenta $\sim 1.4$ to~$1.6\,\mathrm{\AA^{-1}}$ . The pronounced intensity 2.15~eV below the Fermi energy $E_\mathrm{F}$  (defined as $E=0$) derives from  standard one-photon photoemission from the highest occupied molecular orbital (HOMO), whereas the weak intensity 0.25~eV above $E_\mathrm{F}$ originates from two-photon photoemission of the lowest unoccupied molecular orbital (LUMO), populated by the pump pulse before photoionisation. This assignment is unambiguous, because constant energy intensity maps, evaluated at orbital energies $E$ as a function of $k_x$ and $k_y$ and called momentum maps or tomograms in photoemission tomography, are fingerprints of individual orbitals. If the final state of the photoemission process is approximated as a plane wave, the relation is particularly straightforward, because then the momentum maps are closely related to the Fourier transform of the orbital \cite{Puschnig2009} (see Methods). Theoretical momentum maps that have been generated from such transforms are displayed in the excitation scheme Fig.~\ref{Fig:ExpOverview}a for the calculated Kohn-Sham HOMO and LUMO of PTCDA.

In Fig.~\ref{Fig:LongDelays}a, b we present measured momentum maps recorded at the energies of the LUMO and the HOMO, for three different delay times between the pump and probe pulses. Each orbital shows a distinct momentum distribution that can be traced on the ultrafast time scale of the experiment. The fact that the observed patterns are orbital tomograms is confirmed by the concentration of the intensity in rings at parallel wave vector $\simeq 1.4~\mathrm{\AA^{-1}}$ for the HOMO and $\simeq 1.6~\mathrm{\AA^{-1}}$ for the LUMO. The detailed structure of the patterns can be explained from the two molecular orientations that coexist on the surface, labelled  0° and 90° with respect to the laboratory frame of reference (Fig.~\ref{Fig:ExpOverview}a). We expect to observe a superposition of the two corresponding theoretical (Kohn-Sham) momentum maps (Fig.~\ref{Fig:LongDelays}c, d). Moreover, the direction of the incident light breaks the symmetry of the generic theoretical momentum maps. In Fig.~\ref{Fig:LongDelays}c, d, the polarization factor $P(\vec{k}) = |\vec{A} \cdot \vec{k}|^2$ of the probe pulse, incident along the $[1\bar10]$ direction of the Cu substrate, has therefore been included in the theoretical momentum maps. The agreement between theoretical and measured maps is indeed excellent, both for the LUMO around temporal overlap (Figs.~\ref{Fig:LongDelays}a, c) and for the HOMO (Figs.~\ref{Fig:LongDelays}b, d). The fact that the LUMO is not populated for negative delay time (Fig.~\ref{Fig:LongDelays}a, $-46$~fs) confirms that there is no static charge transfer from the Cu(001)-2O surface to the molecules \cite{Yang2018}.

The good agreement between experimental and theoretical momentum maps in Fig.~\ref{Fig:LongDelays}a, c reveals that the measured Dyson orbital in Fig.~\ref{Fig:LongDelays}a resembles the Kohn-Sham LUMO of PTCDA as seen in Fig.~\ref{Fig:ExpOverview}a. This indicates that, to a good approximation, the excited $N$ electron wave function contains the LUMO as a single-electron orbital and that the subsequent photoemission removes the excited electron from the LUMO (Fig.~\ref{Fig:ExpOverview}a). We stress that this intuitive picture of the excitation process in terms of frontier orbitals in the independent electron approximation requires photoemission tomography, and in particular the observation of a good agreement between measured Dyson orbitals and calculated Kohn-Sham orbitals, for its confirmation. In conventional femtosecond spectroscopy, we would be following the evolution of a peak in the energy distribution curve, but would not be able to identify the underlying state as an approximate single-particle orbital, because of the missing momentum space information.

An obvious question addresses the lifetime of the excited state. The single-photon pattern for the HOMO in Fig.~\ref{Fig:LongDelays}b  shows no dependence on the delay time. Obviously, only a very small fraction of the molecules are excited by the pump pulse. In contrast, the integrated intensities in Fig.~\ref{Fig:LongDelays_EDC} reveal that after approximately 100~fs the LUMO population suffers a single-exponential decay with a lifetime of $T_1\simeq 250\,$fs. Importantly, no changes in the excited orbital as such take place during this lifetime, as revealed by its constant pattern in momentum space (Fig.~\ref{Fig:LongDelays}a). For an excited molecule at a metal surface, this longevity is quite remarkable \cite{Zhu2004, Echenique2004}. It not only confirms the potency of the atomically thin CuO layer as an electronic decoupling layer, but also makes future time-resolved investigations of chemical bond-breaking highly promising, since a lifetime of 250 fs exceeds the timescale of typical vibrational motion.
\newline

\noindent{\bf Tracing real-space excitation pathways}

The sensitivity of our tomographic pump-probe experiment to temporal changes in the momentum pattern enables investigations of the excitation mechanism with unprecedented detail, ultimately tracing excitation pathways of electrons in real space. Specifically, we break the symmetry between the two differently oriented PTCDA molecules on the Cu(001)-2O surface and, in a single experiment, follow them separately through the excitation. To this end, we rotate the plane of light incidence relative to the sample by 45° azimuthally as compared to the previous geometry, thus aligning it with the long axis of the 0° molecule (Fig.~\ref{Fig:exc_dynamics}). We find that under these circumstances s-polarised light excites only the 90° molecule (Fig.~\ref{Fig:exc_dynamics}a), while p-polarised light is able to excite both molecules, albeit with surprisingly different excitation dynamics (Fig.~\ref{Fig:exc_dynamics}b): notably, the LUMO pattern of the 0° molecule lights up much earlier and much brighter than the one of the 90° molecule.

A plot of the integrated intensity over the relevant regions in parallel momentum space quantitatively confirms the markedly different behaviour (Fig.~\ref{Fig:exc_dynamics}c): whereas for the 90° molecule the LUMO signal gradually builds up over the duration of the pump pulse and subsequently decays with the time constant of $T_1\simeq 250$~fs, it rises fast for the 0° molecule and reaches a pronounced maximum after $15$~fs, before exhibiting at later times ($t_p >\, \sim 75$~fs) the same slow decay as for the 90° molecule. Upon excitation with s-polarised light, the 90° molecule behaves similar to the 0° molecule under p-polarised excitation (Fig.~\ref{Fig:exc_dynamics}c).

For a conceptual analysis of these findings, we model the experiment by a four-level system consisting of the molecular HOMO $|\phi_1\rangle$ and an occupied metallic state $|\phi_{1'}\rangle$ located below the ultrathin oxide as initial states, the molecular LUMO as intermediate state $|\phi_2\rangle$, and the photoemission final state $|\phi_3\rangle$ (Fig.~\ref{Fig:exc_dynamics}d). On the basis of this model, we explain the differences in the excitation dynamics by two distinct excitation pathways of the electron in the sample before it is photoemitted. In a perturbative description of light-matter interaction, the pump pulse creates a coherent polarisation $|\Psi,t\rangle =  c_{g}(t) e^{-i\omega_gt} |\phi_g\rangle + c_2(t) e^{-i\omega_2t} |\phi_2\rangle$ between ground state $|\phi_g\rangle=|\phi_1\rangle$ or $|\phi_{1'}\rangle$ and excited state $|\phi_2\rangle$ in first order of the electric field. The conversion of this polarisation in second order of the electric field into an excited-state population $n_2$ is governed by phase-destroying elastic scattering processes. In one limit, if this dephasing is fast and the inelastic decay of $|\phi_2\rangle$ is slow, the build-up of $n_2$ follows the time-integrated intensity of the pump pulse. Essentially, this is the situation that we encounter for the 90° molecule with p-polarisation in Fig.~\ref{Fig:exc_dynamics}c. If p-polarised light excites this molecule, the field component $\vec{E}_\perp$ drives a perpendicular electron motion between the substrate and the molecule. The transition then involves an initial state $|\phi_{1'}\rangle$ coupled to a metallic continuum, which results in fast dephasing. While this mechanism in principle applies to both the 0° and 90° molecules (schematic in Fig.~\ref{Fig:exc_dynamics}c), in case of the 0° molecule a competing process takes place (see below), such that the signature of the fast dephasing is only seen in the data for the 90° molecule. To conclude, in this limit the emitted photoelectron can be traced back to the metal.

In the opposite limit, if the dephasing is slow, the Rabi oscillations between ground and excited states driven by the laser field decay slowly. In this coherent regime, the interaction of the electric field of the HHG probe pulse with the polarization $|\Psi,t\rangle$ can contribute to the two-photon photoemission process for short delay times. In addition, if the driving pump laser is slightly detuned from the transition frequency, the excited state can be populated and depopulated by the pump pulse. Both effects give rise to a pronounced peak in the photoemission signal as observed in Fig.~\ref{Fig:exc_dynamics}c for the 90° molecule excited with s-polarised light, as well as for the 0° molecule excited with p-polarised light.

For a quantitative evaluation of the data (see Methods) we have applied a density matrix approach and solved the optical Bloch equation for the four-level system (Fig.~\ref{Fig:exc_dynamics}d). The model clearly confirms the existence of two distinct excitation pathways. We find that the experimental data are described well with an extremely short dephasing time $T_2^{*'}\simeq 3$~fs of the metal state $|\phi_{1'}\rangle$ and a surprisingly long decoherence time $T_2^{12}$ in excess of $150$~fs for the Rabi oscillations between $|\phi_{1}\rangle$ and $|\phi_{2}\rangle$. 

The existence of such a long-lived coherence in an electronic excitation at a metal surface is unexpected \cite{Cui2014,Echenique2004}. It has previously only been observed for image-potential states with high quantum numbers where the electron is mainly located in the vacuum, tens of Angstroms above the surface \cite{Echenique2004}. However, in the present experiment this long coherence time can be rationalised by the fact that the pump pulse induces an in-plane oscillatory electron motion that is confined to the molecule: for both combinations s-polarisation / 90° molecule and p-polarization / 0° molecule, there is an $\vec{E}$ field component along the long axis of the respective molecule (see the sketches in Fig.~\ref{Fig:exc_dynamics}c), i.e., parallel to the HOMO-LUMO transition dipole (see Methods), thus permitting an excitation directly from the HOMO $|\phi_1\rangle$ into the LUMO $|\phi_2\rangle$. If the hybridisation of the molecule with the metal beneath the ultrathin oxide is negligible, which is in fact also revealed by the long inelastic lifetime $T_1$ measured for the LUMO (Fig.~\ref{Fig:LongDelays_EDC}b), this spatial confinement of electron motion in $|\Psi,t\rangle$ to the molecule explains the long coherence.  For this excitation pathway, the emitted photoelectron can clearly be traced back to the HOMO.

It should be noted that for non-resonant excitation, the coherent interaction of pump and probe light can lead to the emission of photoelectrons from the HOMO in a true two-photon process without involving an intermediate state \cite{Echenique2004}. The observed momentum map, which in our experiment is that of the molecular LUMO $|\phi_2\rangle$ and not that of the HOMO $|\phi_1\rangle$, however, provides clear experimental evidence that this is not the case here. The observed peak in the two-photon photoemission signal (Fig.~\ref{Fig:exc_dynamics}c) is thus an unambiguous signature of a persistent coherent HOMO-LUMO polarisation including Rabi oscillations, when the HOMO-LUMO transition is excited by the pump light. This also evidenced by the time shift of the maximum with respect to the cross correlation.
\newline

\noindent{\bf Conclusion}

Time-resolved photoemission tomography allows us to identify and distinguish excitation mechanisms by tracing electrons not only in time, but also in space. This is accomplished by evaluating the orbital momentum maps. Future experiments of this type will allow for studies of molecular electron transfer processes with unprecedented detail. For example, an extension to attosecond time resolution seems feasible, because the availability of momentum space information relaxes pertinent requirements on energy resolution, allowing in turn for an increase in time resolution. Further opportunities arise from the use of terahertz pump pulses in combination with sub-cycle time resolution as demonstrated recently in a different context  \cite{Reimann2018}. We thus anticipate that time-resolved photoemission tomography will soon make it possible to monitor molecular orbitals in real time while chemical bonds are formed or broken. Finally, the observed long decoherence time holds great promise for the applicability of coherent control schemes to manipulate such processes.

\clearpage

\renewcommand{\figurename}{Fig.}
\renewcommand{\thefigure}{\arabic{figure}}

\begin{figure*}
  \centering
  \includegraphics{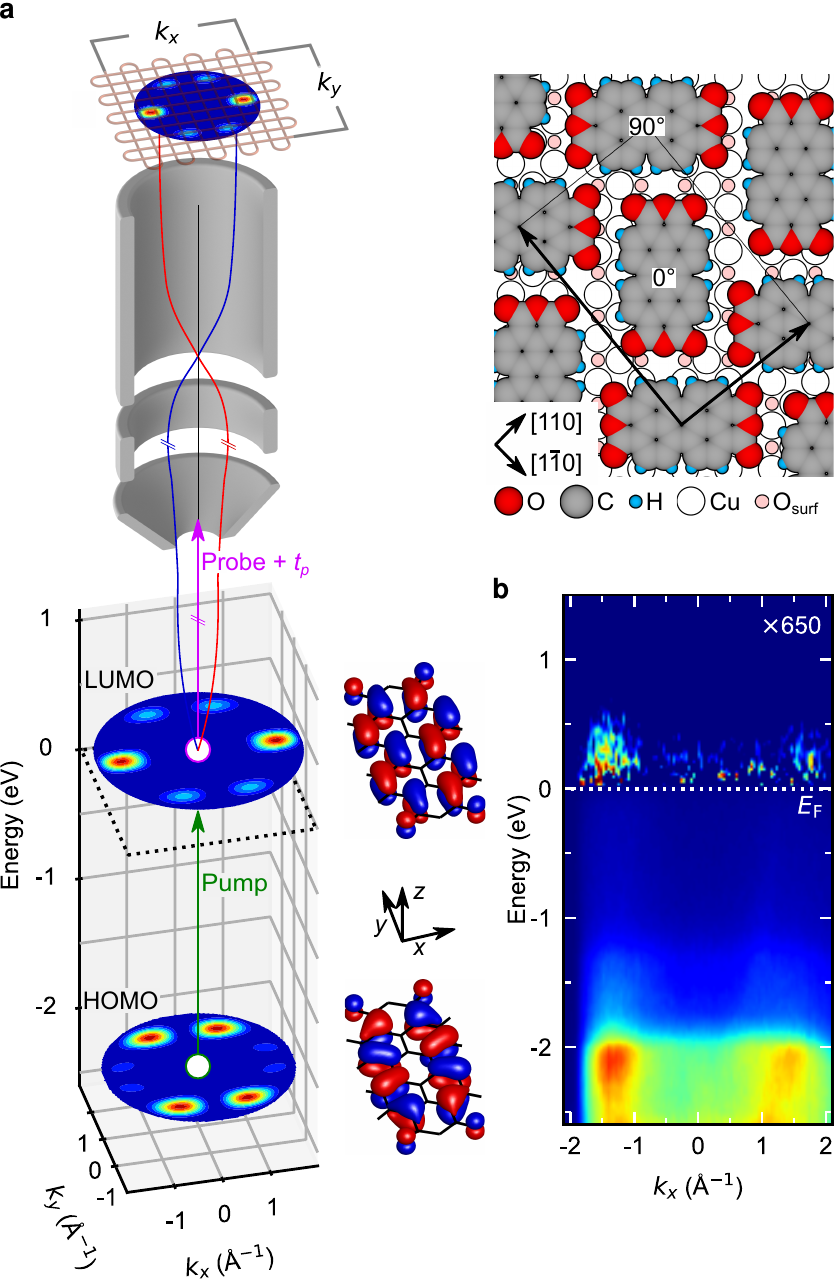}
  \caption{  \setstretch{1.2}
  \textbf{Femtosecond time-resolved photoemission tomography of molecular orbitals.}
  \textbf{a}, Schematic representation of the tomographic pump-probe two-photon photoemission experiment and the time-of-flight momentum microscope. An electron from the HOMO is excited into the LUMO by the pump pulse (green) and then photoemitted from this state by the probe pulse (purple) after a variable delay time $t_p$. 
  The $k_x,\, k_y$ momentum maps indicate the  distribution of photoelectrons that is recorded with the momentum microscope simultaneously with the kinetic energy. The displayed momentum maps, as well as the corresponding real-space orbitals, belong to a molecule with 0° orientation and have been calculated with density functional theory.  The panel also includes a schematic view of a layer of PTCDA on the Cu(001)-2O surface. Each unit mesh of the herringbone structure (indicated by black arrows) contains two molecules, labeled 0° and 90° respectively.
  \textbf{b}, Experimental $I(E,k_x)$ map extracted from a four-dimensional data cube $I(E, k_x, k_y, t_p)$, integrated in the interval $k_y\in[-0.2; 0.2]~\mathrm{\AA^{-1}}$ and plotted around temporal overlap of pump and probe pulses ($t_p = 0$). $E$ is the binding energy. }
   \label{Fig:ExpOverview}
\end{figure*}

\begin{figure*}
  \centering
  \includegraphics[width=\textwidth]{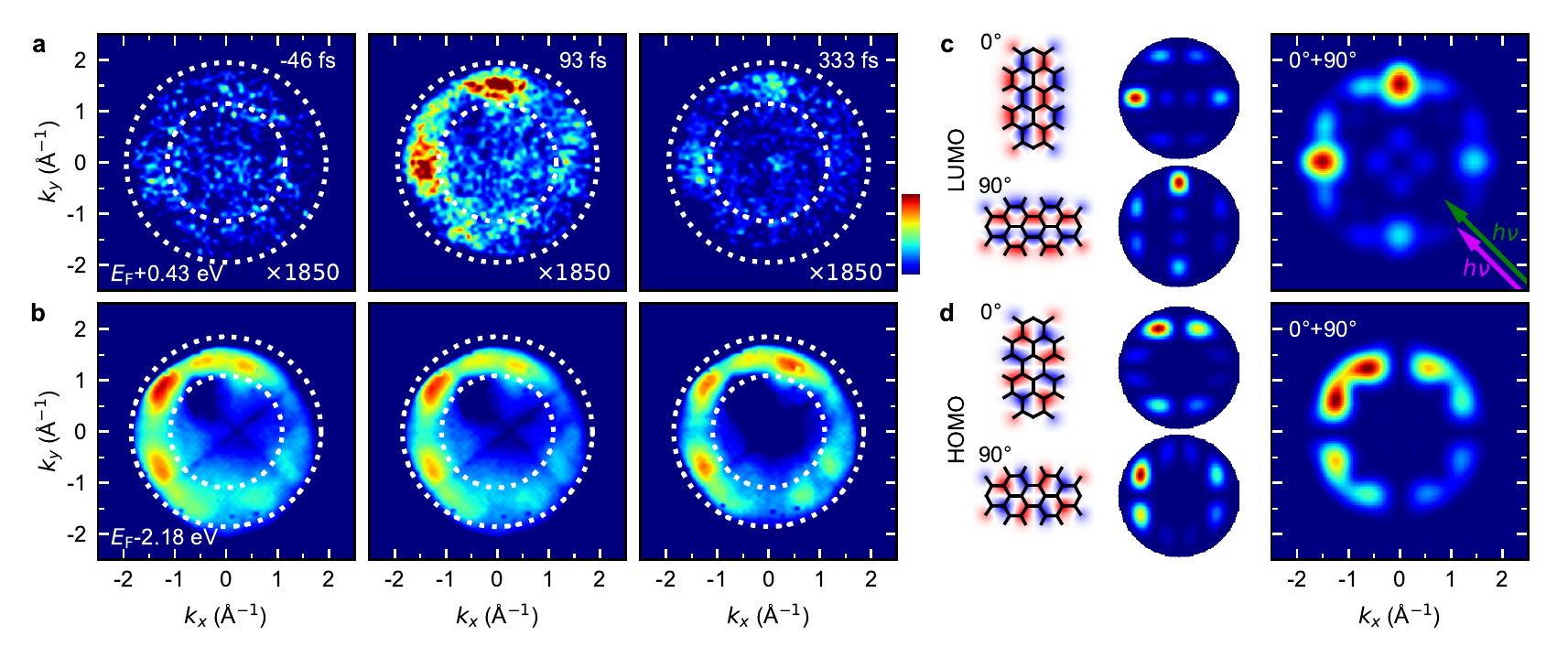}
  \caption{  \setstretch{1.5}
  \textbf{Momentum maps of the frontier orbitals.}  
  \textbf{a}, Experimental LUMO momentum maps obtained from PTCDA on Cu(001)-2O at selected delay times $t_p$. The two dotted circles indicate the momentum integration range used in Fig.~\ref{Fig:LongDelays_EDC}.  \textbf{b}, as \textbf{a}, but for the HOMO.   
  \textbf{c}, Two-dimensional cuts through the LUMO of gas-phase PTCDA, calculated by density functional theory (left), and corresponding theoretical momentum maps of the LUMO, including the polarisation factor $P(\vec{k})$, plotted for the two single orientations 0° and 90° (middle) and their sum (right). \textbf{d}, as \textbf{c}, but for the HOMO. The projected light incidence in panels \textbf{a} to \textbf{d} is 45$^\circ$, as indicated by the two coloured arrows for pump and probe pulses, respectively.
 }
  \label{Fig:LongDelays}
\end{figure*}

\begin{figure}
  \centering
  \includegraphics{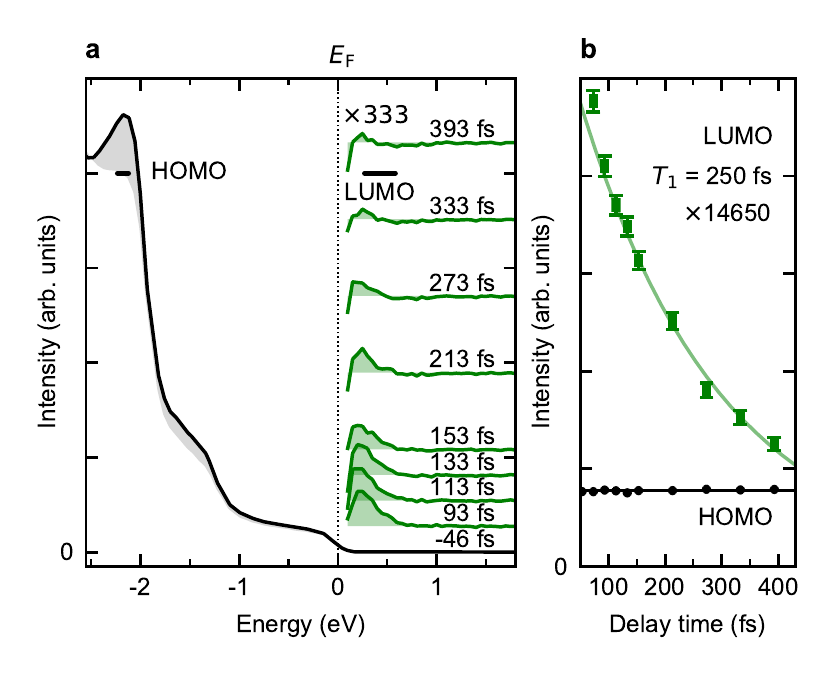}
  \caption{  \setstretch{1.5}
  \textbf{Lifetime of the excited electron in the LUMO.} \textbf{a}, Energy distribution curves of the  experiment in Fig.~\ref{Fig:LongDelays}, with the intensity integrated between the two white dotted circles in Fig.~\ref{Fig:LongDelays}a, b. The black curve belongs to the data obtained for the most negative delay time ($-46$~fs), corresponding to one-photon photoemission from the HOMO and the metal substrate. The grey shaded area indicates the difference to the background, the latter as determined by integrating the intensity within the inner white dotted circle in Fig.~\ref{Fig:LongDelays}a, b. Above $E_\mathrm{F}$, the green curves show the intensity differences between one- and two-photon photoemission as a function of delay time $t_p$. They correspond to photoemission from the LUMO. \textbf{b}, LUMO intensity (green data points with error bars) with fit to a single exponential decay (green line) and HOMO intensity (black data points) with time average (black line). The energy integration ranges used in \textbf{b}, as well as in Figs.~\ref{Fig:LongDelays}a, b, are indicated by thick horizontal bars in \textbf{a}.}
  \label{Fig:LongDelays_EDC}
\end{figure}

\begin{figure*}
  \centering
  \includegraphics[width=\textwidth]{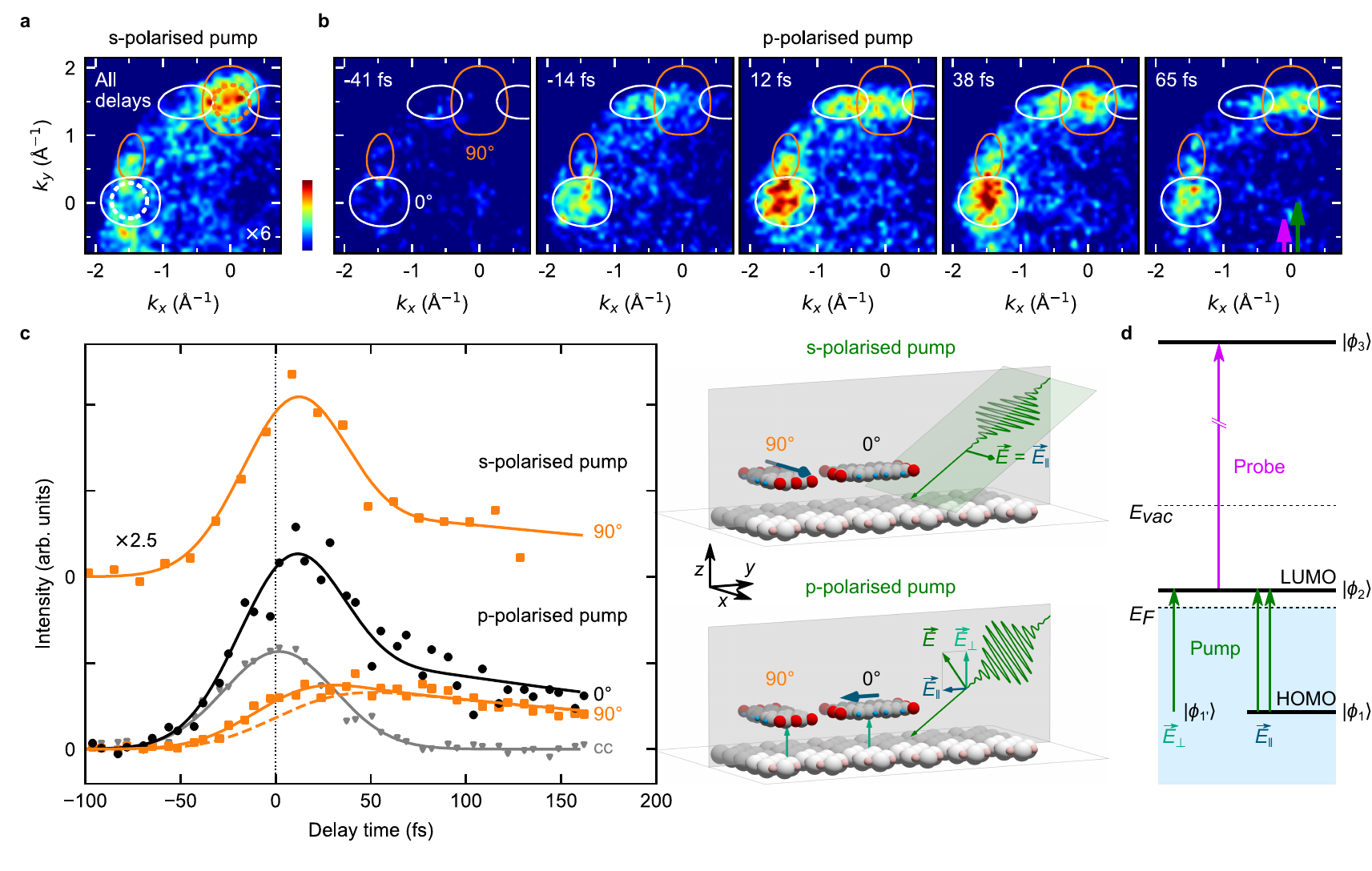}
  \caption{  \setstretch{1.5}
  \textbf{Momentum-resolved LUMO dynamics for different excitation pathways}. \textbf{a}, Momentum map integrated over all delay times for an s-polarised pump pulse with the electric field oscillating in the $x$ direction. The momentum map is overlaid by constant-intensity contours as expected for the 0° (white) and 90° molecules (orange), showing the absence of signal for the 0° molecule.
    \textbf{b}, Momentum maps for a p-polarised pump pulse probed at selected delay times $t_p$ during the rise and fall of the pump pulse. The contours (defined as in \textbf{a}) show that the 0° molecule lights up earlier and stronger than the 90° molecule. 
    \textbf{c}, Temporal evolution of the photoelectron intensity, integrated within the dotted circles in \textbf{a}, of the two distinct LUMOs (0° -- black circles, 90° -- orange squares) for s- and p-polarised pump pulses as indicated. The experimental geometries are illustrated schematically on the right. The solid lines indicate the best fit solution to the optical Bloch equations for the four-level system discussed in the text. The dashed orange line  illustrates a completely incoherent excitation. Grey triangles show the cross-correlation of the pump and probe pulses measured on the sample by recording  photoelectrons at energies far above the excited state $|\phi_2\rangle$.
    \textbf{d}, Schematic representation of the four-level model used to fit the data in \textbf{c}.}
  \label{Fig:exc_dynamics}
\end{figure*}
\clearpage

\newcommand{\Gaertnermatrix}{$\left(\protect\smallmatrix -3.7&~3.0 \\ ~4.8&~5.8 \endsmallmatrix \right)$\xspace}

\newcommand{\bra}[1]{\langle{\phi_{#1}}|}
\newcommand{\ket}[1]{|\phi_{#1}\rangle}

\newpage
\noindent \textbf{Methods}
\newline

\noindent \textbf{Sample preparation and surface structure}

The samples were prepared in a two-step process. The starting point was a Cu(001) surface that had been cleaned beforehand by repeated cycles of Ar$^+$ sputtering (1~keV, incident angle 45°, $30~$min) and annealing (860~K, 30~min). In the first preparation step, this surface was annealed in oxygen atmosphere ($5 \times 10^{-7}~$mbar, 400~K, 30~min), leading to the formation of a Cu(001)-2O missing row reconstruction with the superstructure matrix \Omatrix. This structure, commonly written as Cu(001)-\OWoodN, is formed by oxygen atoms occupying every second hollow site at the surface (which, alone, would form a c$(2 \times 2)$ structure), and removing one quarter of the Cu surface atoms in a way that missing Cu rows in [010] direction emerge \cite{Zeng1989, Wuttig1989, Iddir2007}. Due to the four-fold rotational symmetry of the Cu(001) substrate, two missing row domains occur, as illustrated in Extended Data Fig.~\ref{supp:LEED}a (left vs.\ right side). The formation of these two rotational domains implies that Cu(001)-\OWoodN effectively exhibits $4$mm symmetry, although its unit mesh itself is only $2$mm symmetric.

In the second preparation step, a monolayer film of PTCDA is deposited on Cu(001)-\OWoodN at room temperature. The result is a well-ordered, incommensurate superstructure that is described by the matrix \PTCDAmatrix. The corresponding unit mesh (blue solid rectangle in the upper right quadrant of Extended Data Fig.~\ref{supp:LEED}a) is almost rectangular with $a_1 = 13.38$~\AA, $a_2 = 21.06$~\AA, and $\gamma = 91.0^\circ$. The molecules orient themselves in a herringbone-like pattern, i.e., they adopt an angle of $90^\circ$ with respect to each other, and they are rotated by $\pm 45^\circ$ with respect to the fundamental Cu(001) crystallographic directions $[1 \overline{1} 0$] and [$1 1 0$], an orientation that is confirmed by our photoelectron tomography results.

Due to the effective $4$mm symmetry of Cu(001)-\OWoodN, PTCDA crystallises in four symmetry-equivalent rotational and mirror domains, the structures and unit meshes of which are shown in Extended Data Fig.~\ref{supp:LEED}a. The corresponding measured and simulated LEED patterns are depicted in Extended Data Fig.~\ref{supp:LEED}b and c, respectively, and also reflect the $4$mm symmetry of the system. From a comparison of simulated and measured LEED patterns we estimate the precision of the numbers in the superstructure matrix to approximately $\pm0.1$. Although the system PTCDA on Cu(001)-\OWoodN forms four symmetry-equivalent domains, there are altogether only two different molecular orientations in the entire layer (labelled $0^\circ$ and $90^\circ$ with respect to the laboratory frame), as demonstrated in Extended Data Fig.~\ref{supp:LEED}a. All symmetry operations of the $4$mm symmetric substrate only reproduce these same two orientations.

 The \PTCDAmatrix PTCDA monolayer structure reported here and by Yang et al.~\cite{Yang2018} is exclusively formed when the annealing temperature during the oxidation of the Cu(001) surface (first preparation step) does not exceed $400$~K. It differs from the \Gaertnermatrix structure found by G\"artner et al.~\cite{Gaertner2014} for an oxidation temperature of $470$~K, the essential difference being a $45^\circ$ rotation of the entire PTCDA layer with respect to the substrate. When preparing the oxidised surface at a temperature above $400$~K we observe a coexistence of both structures, caused by an entropy-driven order-disorder phase transition that destroys the long-range order of the missing rows \cite{Duan2010}.

We have verified the structural quality of the sample after each step of the preparation procedure by low energy electron diffraction (LEED) for homogeneity and purity of the phases.
\newline

\noindent \textbf{Experimental setup}

The laser setup for high-harmonic generation is based on the work of Heyl et al.~\cite{Heyl2012} and is depicted schematically in Extended Data Fig.~\ref{extFig:OpticalSetup}. A Ti:sapphire regenerative amplifier, operated at 200~kHz, delivers 800~nm laser pulses with 40~fs duration and 8~µJ pulse energy, which are split 70\% : 30\% into a pump and a probe branch. The pulses in the pump branch are frequency converted in an optical parametric amplifier (OPA) to 540~nm (2.3~eV) with a pulse duration of 50~fs. The pulses in the probe branch are further amplified by a two-pass amplifier and frequency doubled to 400~nm with a pulse duration of 60~fs and pulse energy of 2.5~µJ. For high-harmonic generation these pulses are focused tightly by a 60~mm achromatic lens into a supersonic Krypton jet. The gas jet is produced by injecting pressurised Krypton ($4$~bar) into a UHV chamber ($10^{-8}$~mbar without gas load) through a glass nozzle with a hole diameter of 30~µm. High-harmonic generation at these conditions leads to an almost isolated 7th harmonic (57~nm, 21.7~eV photon energy \cite{Wang2015}) with very high conversion efficiency ($\approx 10^{11} $ photons/s). Collimation and refocusing of the harmonics onto the sample is achieved by two multilayer mirrors which further suppress neighbouring harmonics.

The pump pulses are characterised by a spectrometer and an autocorrelator. They are delayed with respect to the HHG pulses by a linear delay stage and focused by a $f=500$~mm mirror placed outside the vacuum chamber. Inside the vacuum chamber they are redirected by a D-shaped mirror to be almost collinear with the HHG beam. The beam diameter at the sample position is around 100~µm, which leads to fluences on the sample surface of around 50~µJ/cm$^2$. The probe pulse is always p-polarised. The pump pulse can either be p- or s-polarised. For s-polarisation, we rotate the polarisation by introducing an achromatic $\lambda/2$-wave-plate into the pump beam path. Pump and probe pulses are both incident under 70° with respect to the surface normal.

During the measurement, the HHG mirror chamber is separated from the analyser chamber by a 100~nm thin Al-filter which blocks the residual 400~nm beam and prevents contamination of the sample. The vacuum in the analyser chamber is below $2\times 10^{-10}$~mbar, so that we observe no changes in the photoelectron intensity due to surface contaminations over a period of 7 days. In the momentum microscope the photoemitted electrons are projected by a photoemission electron microscope (PEEM)-like lens system onto a time- and position-sensitive delay-line detector \cite{Tusche2016}. The lens system and the detector are separated by a drift tube of 1~m length. The position of electron impact on the detector is proportional to the parallel momentum $(k_x, k_y)$. By measuring the photoelectron intensity $I$ on the detector and the time of flight of the electrons as well as the parallel momentum for each pump-probe delay time $t_p$, we obtain a four-dimensional data set $I(E, k_x, k_y, t_p)$, after converting the kinetic energy $E_\mathrm{kin}$ to the binding energy $E$. The time resolution of our detector is 200~ps, which results in an energy resolution better than 50~meV. The momentum magnification was chosen such that the major features of HOMO and LUMO around 1.4~$\mathrm{\AA^{-1}}$ and 1.6~$\mathrm{\AA^{-1}}$, respectively, are imaged onto the detector with a momentum resolution better than 0.01$\mathrm{\AA^{-1}}$. Low energy electrons are suppressed by applying a retarding field inside the electro-optical system. With this suppression,  we obtain sharp momentum patterns in a range of $\pm2.5$~eV around $E_\mathrm{F}$, sufficient for a simultaneous measurement of LUMO and HOMO.
\newline

\noindent \textbf{Calculation of the momentum maps}

Within the one-step model of photoemission and under the additional assumption that the final state can be approximated by a plane wave, the photoelectron intensity $I_i$ arising from a molecular orbital $i$ is proportional to the product of the modulus square of the Fourier transform  $\tilde{\Psi}_{i} (\vec{k})$ of the initial state wave function $\Psi_{i} (x,y,z)$ and the polarisation factor $P (\vec{k})= |\vec{A} \cdot \vec{k}|^2$ \cite{Puschnig2009},
\begin{equation}
\label{eq:PE1}
I_i(k_x,k_y)  \propto  P (\vec{k}) \left| \tilde{\Psi}_{i} (\vec{k}) \right|^2.
\end{equation}
$\vec{A}$ denotes the vector potential of the incoming photon field (probe pulse). In the Coulomb gauge $\vec{A}$ is parallel to the electric field vector $\vec{E}$. $\vec{k}=(k_x,k_y,k_z)$ is the wave vector of the photoemitted electron, with
\begin{equation}
\label{eq:PE2}
k_z=\sqrt{2mE_\mathrm{kin}/\hbar^2-k_x^2-k_y^2},
\end{equation}
where $E_\mathrm{kin}$ is the kinetic energy of the emitted photoelectron. Geometrically, this last equation corresponds to a hemispherical cut, in  three-dimensional $\vec{k}$ space, by the so-called Ewald sphere with radius $2mE_\mathrm{kin}/\hbar^2$ through the right hand side of Eq.~\ref{eq:PE1}. The maps in Fig.~\ref{Fig:LongDelays}c, d are projections of such cuts into the $k_x,k_y$ plane and can therefore be directly compared to the measured photoelectron intensities $I_i(k_x,k_y)$ as in, e.g., Fig.~\ref{Fig:LongDelays}a, b.  If $P(\vec{k})$ is ignored, as in the momentum maps in Fig.~\ref{Fig:ExpOverview}a, the hemispherical cut is carried out through   $\left| \tilde{\Psi}_{i} (\vec{k}) \right|^2$.

The Fourier transforms $\tilde{\Psi}_{i} (\vec{k})$ of molecular orbitals $\Psi_{i} (x,y,z)$ are computed from density functional theory calculations for a free PTCDA molecule. For this purpose, we utilise NWChem \cite{Valiev2010} and employ a generalised gradient approximation for the exchange-correlation functional \cite{Perdew1996} and the 6-31G* basis set. The presence of the two nonequivalent PTCDA molecules, denoted as $0^\circ$ and $90^\circ$, which are oriented perpendicular to each other, is accounted for by a simple superposition (Fig.~\ref{Fig:LongDelays}c, d). 

When defining the direction of the incident probe laser beam by the two angles $\alpha$ and $\beta$, where $\alpha$ is the incidence angle measured with respect to the surface normal and $\beta$ defines the azimuthal orientation of the incidence plane, the polarization factor is given by
\begin{eqnarray}
P_\mathrm{p} & = & \left| k_x \cos \alpha \cos \beta + k_y \cos \alpha \sin \beta + k_z \sin \alpha \right|^2\\
P_\mathrm{s} & = & \left| -k_x \sin \beta + k_y \cos \beta \right|^2,
\end{eqnarray}
where p and s denote in-plane and out-of-plane polarisations, respectively. 
\newline

\noindent \textbf{Transition dipole of the HOMO-LUMO excitation}

In the gas phase, PTCDA has $D_{2h}$ symmetry and, when aligning the molecule in the $xy$-plane and aligning the long axis along the $y$-direction, the HOMO belongs to the irreducible representation $A_\mathrm{u}$ and the LUMO belongs to $B_\mathrm{2g}$. Assuming a dipole transition, one can group-multiply the irreducible representations of the two states  and the dipole operator to see whether the transition is allowed or not:
\begin{eqnarray}
  \langle \phi_1 | x | \phi_2 \rangle &= A_\mathrm{u} \otimes B_\mathrm{3u} \otimes B_\mathrm{2g}  =& B_\mathrm{1g} \rightarrow  \mathrm{not~allowed}, \\
  \langle \phi_1 | y | \phi_2 \rangle &= A_\mathrm{u} \otimes B_\mathrm{2u} \otimes B_\mathrm{2g}  =& A_g \rightarrow  \mathrm{allowed},\\
  \langle \phi_1 | z | \phi_2 \rangle &= A_\mathrm{u} \otimes B_\mathrm{1u} \otimes B_\mathrm{2g}  =& B_\mathrm{3g} \rightarrow  \mathrm{not~allowed}.
\end{eqnarray}
Hence, the polarisation vector must be aligned along the long molecular axis for the HOMO-LUMO transition to become dipole allowed.
\newline

\noindent \textbf{Density matrix calculations for the four-level model}

For the theoretical description of the data of Fig.~\ref{Fig:exc_dynamics} we apply the usual phenomenological model of two-photon photoemission \cite{Klamroth01} and consider an initial state $\ket{1}$ (HOMO), an intermediate state $\ket{2}$ (LUMO) and a photoemission final state $\ket{3}$. They are coupled to the time-dependent electric fields $\vec{E}_a(t)=\mathcal{E}_a(t)\vec{e}_ae^{i\omega_at}+\mathrm{c.c.}$ and $\vec{E}_b(t-t_p)=\mathcal{E}_b(t-t_p)\vec{e}_be^{i\omega_b(t-t_p)}+\mathrm{c.c.}$ of the pump and probe pulses by dipole interactions. $\mathcal{E}_a(t)$ and $\mathcal{E}_b(t)$ are the pulse envelopes, $\vec{e}_a$, $\vec{e}_b$ the polarisation vectors, and $\omega_a$, $\omega_b$ the carrier frequencies of the pump and probe pulses, respectively. $t_p$ denotes the delay time of the probe pulse with respect to the pump pulse. In order to allow for the excitation of the LUMO from the metal substrate we add an additional initial state $\ket{1'}$ such that the Hamiltonian of the system reads
\begin{equation}
\begin{split}
H=\sum_{i=1,1',2,3} \epsilon_i\ket{i}\bra{i} - \ket{2}\bra{1}\mu_{21}\vec{E}_a(t) \\ - \ket{2}\bra{1'}\mu_{21'}\vec{E}_a(t) - \ket{3}\bra{2}\mu_{32}\vec{E}_b(t-t_p)
\label{eq:Hamiltonian}
\end{split}
\end{equation}
Here, the energies $\epsilon_1$ and $\epsilon_2$ are the measured HOMO and LUMO energies, $\epsilon_3$ is the energy of the detected photoelectron, and $\mu_{ij}$ are the transition matrix elements. We only consider optical transitions from $\ket{1}$ and $\ket{1'}$ to $\ket{2}$ by the pump pulse and from $\ket{2}$ to $\ket{3}$ by the probe pulse, which is well justified by the large difference between their photon energies. Relaxation is treated by considering an ensemble average of states $\ket{i}$ coupled to a bath. The time evolution of the system is then described by the Liouville-von-Neumann equation for the density operator $\rho$ 
\begin{equation}
\partial_t\rho = \frac{1}{i \hbar}[H,\rho] + \partial_t\rho^\mathrm{relax}.
\label{eq:LvN}
\end{equation}
The density operator is defined as $\rho= \sum_i p_i \ket{i} \bra{i}$, where $p_i$ is the probability to find the system in state $i$. The diagonal elements $\rho_{ii}(t)$ represent the time-dependent population $n_i(t)$ of the states $\ket{i}$, the off-diagonal elements $\rho_{ij}(t)$ are optically induced coherences between states $\ket{i}$ and $\ket{j}$. The diagonal relaxation terms $\rho_{ii}^\mathrm{relax}$ describe energy relaxation due to inelastic scattering. We only consider this term for the intermediate state $\ket{2}$. In the Markov approximation it becomes $\partial_t \rho_{22}^\mathrm{relax} = -\rho_{22}/T_1$ and describes an exponential decay of $n_2(t)$. The off-diagonal terms of $\rho^\mathrm{relax}$ describe dephasing between $\ket{i}$ and $\ket{j}$ with a time-constant $T_2^{ij}$ ($\partial_t\rho_{ij}^\mathrm{relax} = -\rho_{ij}/T_2^{ij}$). $T_2^{ij}$ generally has contributions due to elastic scattering, so-called true dephasing, in addition to dephasing caused simply by population decay due to the inelastic $T_1$. In our model we assume similar couplings of HOMO and LUMO to the metal and describe them by the same true dephasing time $T_2^*$. Then we obtain $1/T_2^{12} = 1/2T_1 + 2/T_2^*$ and $1/T_2^{1'2}=1/2T_1 + 1/T_2^{*'} + 1/T_{2}^*$ with the true dephasing of the metallic state $\ket{1'}$ denoted by $T_2^{*'}$. Dephasing in the final photoelectron state $\ket{3}$ is neglected in our model, which results in $1/T_2^{23} = 1/2T_1 + 1/T_2^*$, $T_2^{13} = T_2^*$ and $T_2^{1'3} = T_2^{*'}$.

Eq.~(\ref{eq:LvN}) can be solved by perturbative expansions into powers of the electric fields $\vec{E}_a(t)$ and $\vec{E}_b(t)$. The measured photoelectron intensity $I$ is described by $\rho_{33}(t\rightarrow\infty)$. It is obtained by a fourth-order expansion which results in a set of nine coupled linear differential equations. These optical Bloch equations are solved numerically by applying the rotating frame approximation which eliminates fast variations with the carrier frequencies $\omega_a$ and $\omega_b$. This approximation is good for pulse envelopes of pump and probe pulses $\mathcal{E}_a(t)$ and $\mathcal{E}_b(t)$ that vary slowly compared to the respective carrier frequencies $\omega_a$ and $\omega_b$. It is justified in our experiment since both $\vec{E}_a(t)$ of the visible pump and $\vec{E}_b(t)$ of the HHG probe go through several tens of optical cycles within the half width of 65~fs and 20~fs of the respective Gaussian envelopes.

Before the interaction with the laser field $(t\rightarrow-\infty)$ the system is in its ground state, with all $\rho_{ij} = 0$, except for $\rho_{11} = 1$ and $\rho_{1'1'} = 1$. There are different pathways that connect the initial states $\ket{1}$ and $\ket{1'}$ with the photoemission final state $\ket{3}$. Long dephasing times and off-resonant excitation can lead to interference between these pathways and to a dependence of the photoemission intensity $\rho_{33}(t\rightarrow\infty)$ on the delay time $t_p$ which is much faster than the intermediate state lifetime $T_1$.

The calculated intensity traces as a function of the pump-probe delay time $t_p$ are fitted to the experimental data using $T_2^*$ and $T_2^{*'}$ and the overall intensity as free parameters. The lifetime $T_1$ is set to the 250~fs determined experimentally for long delay times (Fig.~\ref{Fig:LongDelays_EDC}b). Also, the energies $\epsilon_1$ and $\epsilon_2$ and the carrier frequencies $\omega_a$ and $\omega_b$ are taken from the experiment. This results in a slight detuning of 47 meV between the HOMO-LUMO transition $\epsilon_2-\epsilon_1$ and pump photon energy $\hbar\omega_a$. The final state energy is $\epsilon_3=\epsilon_2+\hbar\omega_b$, and for the energy of the metal state we assume $\epsilon_{1'}=\epsilon_2-\hbar\omega_a$. The magnitudes of the dipole matrix elements $\mu_{ij}$ are chosen such that no significant depletion of the ground state by the pump pulse, nor that of the excited state by the probe pulse, occur. For the excitation of the $90^\circ$ molecule by s(p)-polarised pump light, the matrix elements $\mu_{21'}$ ($\mu_{21}$) are set to zero, respectively. For the excitation of the $0^\circ$ molecule by p-polarised pump light the ratio $\mu_{21'} / \mu_{21}$ is set such that the population transfer from the metallic initial state $\ket{1'}$ into the LUMO $\ket{2}$ is the same as for the $90^\circ$ molecule.

As a result of the fitting procedure we find a true dephasing time $T_2^{*} = 490$~fs,  which results in a long decoherence time of the HOMO-LUMO polarisation of $T_2^{12} = 164$~fs. In contrast, the pump-induced coherence between the metallic ground state $\ket{1'}$ and the LUMO $\ket{2}$ dephases very fast, $T_2^{1'2} \simeq T_{2}^{*'} = 3$~fs.

\newpage

\begin{acknowledgments}

\noindent \textbf{Acknowledgements}  R.W., U.H., F.C.B., C.K., and F.S.T. acknowledge financial support by the Deutsche Forschungsgemeinschaft (DFG, German Research Foundation), Project-ID 223848855-SFB 1083. P.P. acknowledges financial support from the Austrian Science Fund (FWF), project I3731. S.S, F.C.B., and F.S.T. acknowledge financial support from the DFG, project Po 2226/2-1. We thank A. Haags for fruitful discussion.
\newline

\noindent \textbf{Author contributions}
C.K., F.S.T., and U.H. conceived the research.
X.Y., S.S., C.K., and F.S.T. chose the sample.
M.R. and F.C.B. prepared the samples and C.K. analysed its structure.
R.W., M.R., K.S., L.M., and J.G. performed the experiment.
R.W., K.S., C.K., F.C.B., F.S.T., and U.H. analysed the data and developed the physical model.
D.B., P.P., and S.S. simulated the momentum maps of the PTCDA molecule in the gas phase.
R.W., C.K., and F.C.B. prepared the figures.
R.W., C.K., F.C.B., F.S.T., and U.H. wrote the paper.
\newline

\noindent \textbf{Correspondence and requests for materials} should be addressed to R.W.~(email: wallauer@staff.uni-marburg.de) and F.S.T.~(email: s.tautz@fz-juelich.de).
\newline

\end{acknowledgments}

\newcommand{\quarter}{$(\frac{1}{4} ~ \frac{1}{4})$\xspace}

\renewcommand{\figurename}{Extended Data Fig.}
\renewcommand{\thefigure}{\arabic{figure}}
\setcounter{figure}{0}

  \begin{figure*}[ht]
    \centering
    \includegraphics[width=\textwidth]{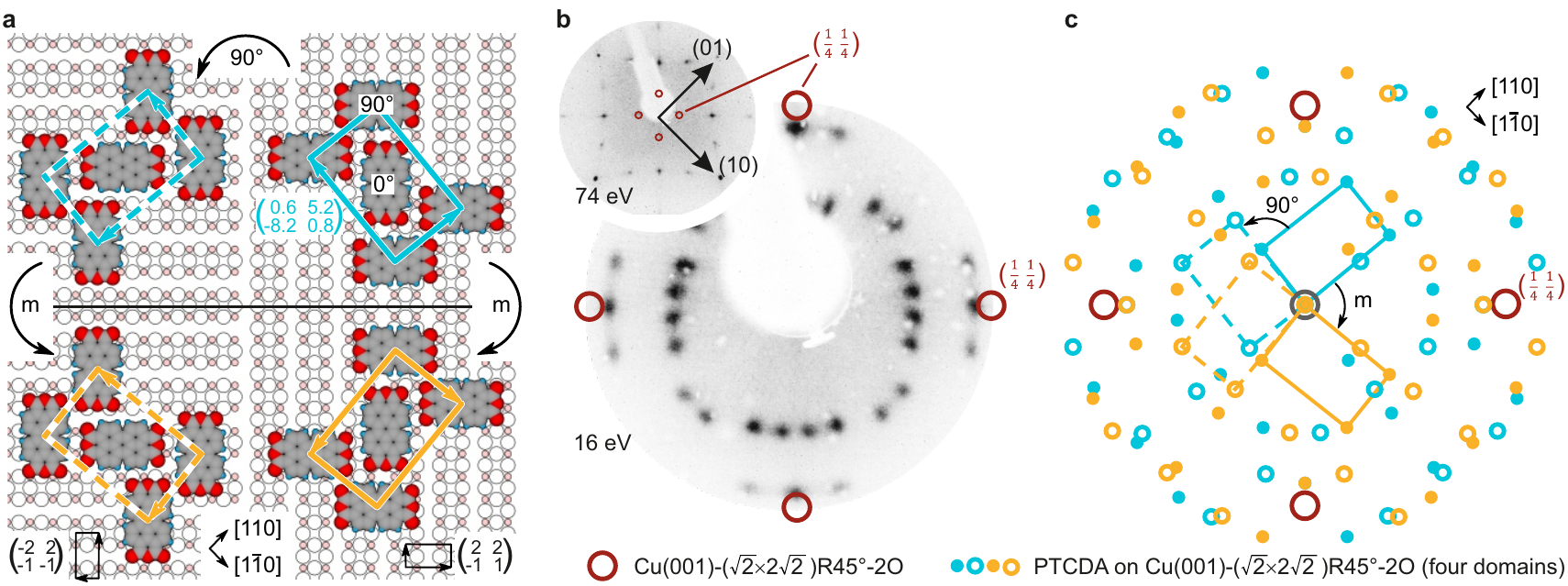}   
    \caption{  \setstretch{1.5}
    \textbf{Structure of PTCDA on the Cu(001)-\OWoodN surface.}
    \textbf{a}, Adsorption geometry of the PTCDA \PTCDAmatrix structure on Cu(001)-\OWoodN. All four equivalent domains resulting from the effective $4$mm symmetry of the Cu(001)-\OWoodN surface are shown. The unit mesh is drawn with solid blue lines, its rotational domain ($90^\circ$ counterclockwise) as dashed lines (upper right and left, respectively). The corresponding mirror domains are shown in the bottom part with orange unit meshes. 
    \textbf{b}, Measured microchannel plate LEED pattern, recorded with 16~eV electron energy. All visible spots arise from the PTCDA layer. The positions of the \quarter spots originating from the Cu(001)-\OWoodN reconstruction are marked by dark red circles. Because of the low electron energy, these LEED spots are hardly visible, unlike in the inset, which was recorded with an electron energy of $74$~eV from the same sample before PTCDA deposition. 
    \textbf{c}, Simulated LEED pattern of the PTCDA overlayer. The LEED spots and reciprocal unit meshes from all four domains are shown (colour coding as in \textbf{a}, full and open circles correspond to solid and dashed lines, respectively). }
    \label{supp:LEED}
  \end{figure*}

  \begin{figure*}
  	\centering
  	\includegraphics[width=\textwidth]{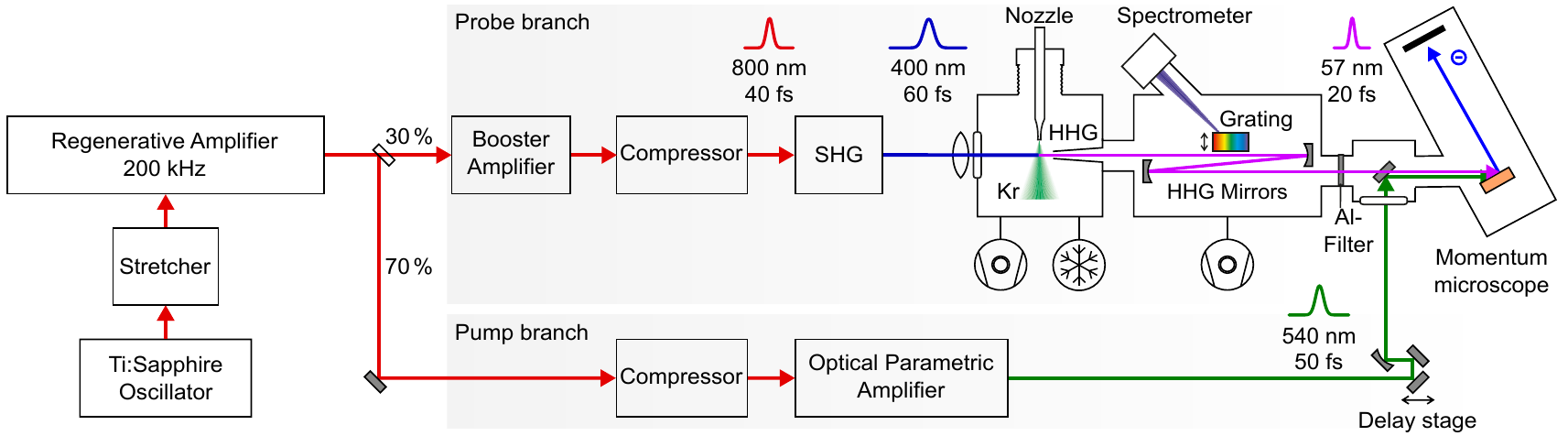}
  	\caption{  \setstretch{1.5} \textbf{Overview of the optical setup.} The regenerative amplifier output  is divided into a pump and a probe branch with an intensity ratio of 70\% : 30\%. After compression, visible pump pulses are provided by an optical parametric amplifier operated at 540~nm. The probe beam is further amplified in a two-pass booster amplifier and compressed independently of the pump pulse. Before entering the high-harmonic generation (HHG) chamber, the 800~nm probe pulses are frequency-doubled (second harmonic generation, SHG). Two multilayer mirrors collimate and refocus the high-harmonic beam onto the sample and suppress neighbouring harmonics. The pump beam is focused by a spherical mirror placed outside the UHV chamber and is directed inside the chamber by a D-shaped mirror in order to overlap almost collinearly with the HHG beam on the sample.}
  	\label{extFig:OpticalSetup}
  \end{figure*}

\end{document}